\documentclass[11pt]{article}

\usepackage{amssymb,amsmath,epsfig,float,times} 
\newtheorem{theorem}{Theorem}
\newtheorem{lemma}[theorem]{Lemma}
\newtheorem{proposition}[theorem]{Proposition}

\newcounter{Remark}

\newcounter{Example}

\newcommand{\theremark}{{\arabic{Remark}}}
\newenvironment{remark}
{\refstepcounter{Remark}
{\noindent{\sc Remark} \theremark:}
{}}

\newcommand{\theexample}{{\arabic{Example}}}
\evensidemargin \oddsidemargin
\newcommand{\ol}{\setlength{\itemsep}{-40pt}\begin{enumerate}}
\newcommand{\eol}{\end{enumerate}\setlength{\itemsep}{-\parsep}}
\newcommand{\ignore}[1]{}
\newcommand\proof{\noindent{\sc Proof. }}
\newcommand\qed{\hfill \rule{1.2mm}{2.8mm}}

\newcommand\bysame{\rule[1mm]{1cm}{.025cm}}

\def\supp{\qopname\relax o{supp}}
\def\dist{\qopname\relax o{d}}
\def\arccot{\qopname\relax o{arccot}}

\newcommand\dgv{{\delta_{\text{\rm GV}}}}

\newcommand\nc\newcommand
\def\Con{\qopname\relax o{Con}}
\newcommand{\logasymp}{\cong}

\nc\bfa{{\bf a}}\nc\bfA{{\bf A}}\nc\cA{{\mathcal A}}
\nc\bfb{{\bf b}}\nc\bfB{{\bf B}}\nc\cB{{\mathcal B}}
\nc\bfc{{\bf c}}\nc\bfC{{\bf C}}\nc\cC{{\mathcal C}}
\nc\bfd{{\bf d}}\nc\bfD{{\bf D}}\nc\cD{{\mathcal D}}
\nc\bfe{{\bf e}}\nc\bfE{{\bf E}}\nc\cE{{\mathcal E}}
\nc\bff{{\bf f}}\nc\bfF{{\bf F}}\nc\cF{{\mathcal F}}
\nc\bfg{{\bf g}}\nc\bfG{{\bf G}}\nc\cG{{\mathcal G}}
\nc\bfh{{\bf h}}\nc\bfH{{\bf H}}\nc\cH{{\mathcal H}}
\nc\bfi{{\bf i}}\nc\bfI{{\bf I}}\nc\cI{{\mathcal I}}
\nc\bfj{{\bf j}}\nc\bfJ{{\bf J}}\nc\cJ{{\mathcal J}}
\nc\bfk{{\bf k}}\nc\bfK{{\bf K}}\nc\cK{{\mathcal K}}
\nc\bfl{{\bf l}}\nc\bfL{{\bf L}}\nc\cL{{\mathcal L}}
\nc\bfm{{\bf m}}\nc\bfM{{\bf M}}\nc\cM{{\mathcal M}}
\nc\bfn{{\bf n}}\nc\bfN{{\bf N}}\nc\cN{{\mathcal N}}
\nc\bfo{{\bf o}}\nc\bfO{{\bf O}}\nc\cO{{\mathcal O}}
\nc\bfp{{\bf p}}\nc\bfP{{\bf P}}\nc\cP{{\mathcal P}}
\nc\bfq{{\bf q}}\nc\bfQ{{\bf Q}}\nc\cQ{{\mathcal Q}}
\nc\bfr{{\bf r}}\nc\bfR{{\bf R}}\nc\cR{{\mathcal R}}
\nc\bfs{{\bf s}}\nc\bfS{{\bf S}}\nc\cS{{\mathcal S}}
\nc\bft{{\bf t}}\nc\bfT{{\bf T}}\nc\cT{{\mathcal T}}
\nc\bfu{{\bf u}}\nc\bfU{{\bf U}}\nc\cU{{\mathcal U}}
\nc\bfv{{\bf v}}\nc\bfV{{\bf V}}\nc\cV{{\mathcal V}}
\nc\bfw{{\bf w}}\nc\bfW{{\bf W}}\nc\cW{{\mathcal W}}
\nc\bfx{{\bf x}}\nc\bfX{{\bf Z}}\nc\cX{{\mathcal X}}
\nc\bfy{{\bf y}}\nc\bfY{{\bf Y}}\nc\cY{{\mathcal Y}}
\nc\bfz{{\bf z}}\nc\bfZ{{\bf Z}}\nc\cZ{{\mathcal Z}}
\nc\od{{\bar d}}\nc\ow{{\bar w}}\nc\odelta{{\bar\delta}}
\nc\ox{{\bar x}}\nc\oy{{\bar y}}\nc\ou{{\bar u}}
\newcommand{\remove}[1]{}

\nc\gb{{\mbox{\german{b}}}}
\newcommand\reals{{\mathbb R}}

\begin{document}

\title{Improved error bounds for the erasure/list scheme: the binary
and spherical cases}

\author{Alexander Barg
\thanks{University of Maryland, College Park, MD 20742, 
e-mail: abarg@ieee.org.
Supported in part by NSF grant CCR-0310961. }
}
\date{}

\maketitle

\begin{abstract}
We derive improved bounds on the error and erasure 
rate for spherical codes and for binary linear codes 
under Forney's erasure/list decoding scheme and prove some related results. 
\end{abstract}


\section{Introduction}\label{sec:intro}

The subject of error bounds for various decoding schemes has been a
central topic in information theory in its first decades. With the
success of turbo codes and other iterative decoding schemes this subject
again became the focus of continued attention through the last decade.
In the early days the major effort in deriving error bounds went into
establishing the best attainable error exponents (for instance, 
Shannon's reliability function of channels). This approach is
reflected in most textbooks on information theory that deal with this
subject \cite{gal63,gal68,vit79,bla87,csi81}. 
Lately the attention has shifted from considering average
properties of code ensembles to bounding the error probability of
decoding of a particular code whose distance distribution is known
or can be estimated. Focusing on a particular code instead of
an ensemble of codes makes it possible to analyze the error
probability by a geometric approach rather than Chernoff bounds.
These studies gained momentum after influential 
research of G. Poltyrev in \cite{pol94a,pol94b}; 
see \cite{sha02} and references therein. 

It is interesting to note that Shannon \cite{sha59}
also relied on a geometric
derivation in his paper on the error bounds for spherical codes and
the Gaussian channel. 
The starting point of the present research was an attempt to derive
Shannon's results via the distance
distribution of the code (recall that about the
original derivation the author wrote: ``It might be said that the
algebra involved is in several places unusually tedious''). 
It turns out that in this way the results
of \cite{sha59} can be obtained by a simpler, more intuitive argument.
To add a new element to this study, we consider a version of
Forney's erasure/list decoding scheme \cite{for66},\,\cite{for68b}.
To define it, let $C$ be a code in a metric (observation) space $X$ with 
the metric $\dist(\cdot,\cdot)$ and let $t\ge 0.$ 
The decoding function $\psi_t$ is defined as follows:
$\psi_t(\bfy)=\bfx$ if for all code vectors
$\bfx'\neq\bfx$ the distance $\dist(\bfx',\bfy)-\dist(\bfx,\bfy)\ge 2t.$
For all other points in $X$ the decoding result is undefined and will be 
called erasure below.

We will be interested in the best attainable exponents of error
and erasure probabilities, denoted $E_e$ and $E_x$ below.
Error bounds for this decoding for general discrete memoryless
channels were derived in \cite{for66},\,\cite{for68b}. In particular,
they imply bounds on $E_e$ and $E_x$ for unrestricted codes in
the Hamming space used over a binary symmetric channel. 
The case of linear codes was addressed by Blokh and
Zyablov \cite{blo82}. Error bounds for this decoding
method in the case of spherical codes are not available in the literature.

The text is organized as follows. In Sect. \ref{sect:Hamming} we
address the technically easier and more familiar case of binary 
linear codes. The main goal of this part is to develop geometric
intuition in a more familiar situation and then to rely on it in a more 
difficult case of spherical codes. However, as a byproduct, we
obtain an improvement of the bounds of \cite{blo82} on $E_e$ and $E_x$. 
Moreover, the method we use 
is arguably easier to understand than the results in \cite{blo82}. 

In Sect. \ref{sect:rH} we consider the case of bounded distance decoding 
and some other related questions.

In the second part we study spherical codes. For the Gaussian channel we 
obtain a pair of bounds that specifies the trade-off between the error 
and erasure 
events. For $t=0$ the bounds reduce to Shannon's lower bound on the 
error exponent of maximum likelihood decoding.
In our calculations we rely on the distance distribution of codes.
Note that Shannon's derivation \cite{sha59}, although geometric in nature, 
takes a somewhat different route, performing averaging 
of the error probability over the choice of codes. This method is not the
best known for low noise because average codes contain small
distances, so expurgation of the code ensemble is needed to
obtain a good bound for low rates. In contrast, we begin with choosing 
codes with large minimum distance and obtain the complete result
by a single argument. Since we operate in terms of the distance 
distribution, we will obtain some new insights into the decoding geometry 
of spherical codes in the course of our derivation. 
We also outline a derivation
of Shannon's error bounds \cite{sha59} by an approach
which is arguably simpler than both the original proof and Gallager's
proof in \cite{gal68}. 
The proof method considered exhibits a close 
analogy between spherical codes and codes in $\{0,1\}^n$ if one makes 
allowance for some peculiarities of discrete geometry. 

We also derive error bounds for bounded distance decoding 
of spherical codes. This problem was mentioned in \cite{wyn65},
however the focus of that paper is on different questions.
In particular, we address the question of the probability of undetected
error with spherical codes, in the sense specified in the main text,
and establish the asymptotic behavior of this quantity.

\section{The binary case}\label{sect:Hamming}

Let $X=\{0,1\}^n$ be the binary Hamming space with distance 
$\dist(\cdot,\cdot)$. We consider linear
codes $C\subset X$ of rate $R=n^{-1}\log_2 |C|$ used over
a binary symmetric channel with crossover probability 
$p\in (0,1/2).$ 

For a code $C\subset X$
consider a decoding mapping mapping $\psi_t:X\to C$ defined as follows: 
$\psi_t(\bfx)=\bfc$ if for all code vectors
$\bfc'\neq\bfc$ the distance $\dist(\bfc',\bfx)-\dist(\bfc,\bfx)\ge 2t$
for some nonnegative integer $t=\tau n$. 
For all other points in $X$ the decoding
result is undefined, and will be called erasure below. For the case of 
complete decoding we write $\psi$ instead of $\psi_0.$

Let us introduce notation. 
Denote by $A_w,w=0,1,\dots,n$ the weight distribution
of $C$. For a code of minimum distance $d$ we have $A_0=1$, 
$A_1=\dots=A_{d-1}=0$. Let us introduce the {\em weight profile} of the code:
for $\omega=w/n, w=0,1,\dots, n$ let
\[
\alpha(\omega)=\frac 1n \log_2 A_{w},
\]
where $\log 0=-\infty.$

Let $\cA_0=1,\cA_w=\lfloor\binom nw 2^{-n(1-R)}\rfloor,w=1,\dots,n$, 
\[
T(x,y)=-x\log_2 y-(1-x)\log_2(1-y),
\]
$h(x):=T(x,x), D(x\|y):=T(x,y)-h(x)$. Throughout the rest of the text
$\dgv(R)=h^{-1}(1-R)$ is the relative Gilbert-Varshamov (GV) distance and 
$d=d_{\text{GV}}=\lfloor\dgv(R) n\rfloor.$ 
\remove{Let
\[
\rho(\lambda)=\frac{p^{1/(1+\lambda)}}{p^{1/(1+\lambda)}+
(1-p)^{1/(1+\lambda)}},
\]
$R_{\lambda}=1-h(\rho(\lambda))$. Letting $\lambda=1$ and $\rho_0:=\rho(1)$
we obtain the ``critical rate'' of the channel, denoted below by $R_c.$ }
Let $E_0(R,p)$ be the Gallager bound on the reliability function
of the channel \cite[pp.34-36]{gal63}:
\begin{equation}\label{eq:Gallager}
E_0(R,p)=\left\{\begin{array}{l@{\quad}l@{\qquad}c}
-\dgv\log_2 2\sqrt {p(1-p)} &0\le R\le R_e, &{\rm (a)}\\[2mm]
D(\rho_0\|p)+R_c-R  
&R_e \le R\le R_{c} ,  &{\rm (b)}\\[2mm]
D(\dgv(R)\|p), &R_c\le R \le 1-h(p), &{\rm (c)}
\end{array}\right.
\end{equation}
where 
\[
\rho_0=\frac{\sqrt p}{\sqrt p+\sqrt{1-p}}, \quad\omega_0=2\rho_0(1-\rho_0),
\]
\[
R_{e}=1-h(\omega_0), \quad R_c=1-h(\rho_0).
\]
Denote by $S_r(0)$ a ball of radius $r$ in $X$ with center at $0$ and by
\[
p_{i,j}^k=|\{\bfz\in X: \dist(\bfz,\bfx)=i,\dist(\bfz,\bfy)=j;
\dist(\bfx,\bfy)=k\|
\]
the number of triangles in $X$ with a fixed side of length $k.$
Let $\nu=\log_2((1-p)/p).$

\remove{Let $E_e(R,p,\tau)$ and $E_x(R,p,\tau)$ be the best attainable error 
and erasure exponents for binary codes under $\psi_t,$ where
$t=\tau n.$ By a result of Forney \cite{for68b}, \cite{csi81} there exist 
binary codes of rate $R$ and sufficiently large length $n$}

For unrestricted codes various lower bounds on the exponents $E_e,E_x$
were given in Forney \cite{for68b}.
For linear binary codes the following theorem was proved by Blokh and
Zyablov.

\begin{theorem}\label{thm:bz} {\rm\cite{blo82}} 
For $0\le R <R_c$
\begin{eqnarray}
E_e(R,p,\tau)&\ge& E_0(R,p)+\nu \tau \label{eq:Ee-low}\\
E_x(R,p,\tau)&\ge& E_0(R,p)-\nu \tau,\label{eq:Ex-low}
\end{eqnarray}
For $R\ge R_c$
\begin{eqnarray}
E_e(R,p,\tau)&\ge& E_0(R,p)+2\tau D'(\delta\|p)|_{\delta=\dgv(R)}
\label{eq:Ee-high}\\
E_x(R,p,\tau )&\ge& E_0(R,p)-2\tau D'(\delta\|p)|_{\delta=\dgv(R)} .
\label{eq:Ex-high}
\end{eqnarray}
\end{theorem}

Note that the case $t=0$ corresponds to maximum likelihood decoding,
and the bound on $E_e$ turns into $E_0$. Erasure rate in this case is
of course zero though (\ref{eq:Ex-low}), (\ref{eq:Ex-high}) give a
positive value, because by the nature of the argument the erasure 
probability $P_x$ is estimated by the sum $P_e+P_x$.

\begin{remark}\label{remark}
Note also that by (\ref{eq:Ex-low}), the exponent $E_x=0$ for rates in the
range close to the channel capacity. In this range the value of the
undetected error exponent $E_e$ can be claimed arbitrarily large if we modify
the decoding function to claim an erasure for all transmissions. Thus,
in effect Theorem \ref{thm:bz} contains a nontrivial claim {\em 
only for those values of the code rate $R$ for which $E_x>0$}, i.e.,
for which the right-hand side of (\ref{eq:Ex-high}) is positive. Of course,
even when $E_x=0$ i.e. decoding results in erasures in almost all
transmissions, it is still useful to know how often we will run into
an undetected error.
\end{remark}

The aim of this section is to derive lower bounds on the exponents
which are better than the estimates (\ref{eq:Ee-low})-(\ref{eq:Ex-high})
for most values of $R>0$.  
To state the results we need the following definitions: $u=p(1-p),$ 
\begin{align*}
\rho_0^{\pm}&=\frac{\sqrt {u+\tau^2(1-2p)^2}-p(1\pm 2\tau)\pm\tau}{1-2p}\\
\omega_0&=2\rho_0^{\pm}(1-\rho_0^{\pm})\pm 2\tau(1-2\rho_0^\pm)\\
&=2\frac{\sqrt{u+\tau^2(1-4u)}-2u}{1-4u}
\end{align*}
\begin{equation*}
M_\pm=\left\{\begin{array}{l@{\quad}l@{\qquad}c}
-\dgv(R)(h(\frac12+\frac\tau\dgv)+\frac12\log_2 u)\pm\nu\tau
&0\le R\le 1-h(\omega_0) &{\rm (a)}\\[2mm]
D(\rho_0^\pm\|p)+1-R-h(\rho_0^\pm\mp 2\tau) &1-h(\omega_0)< R\le 
1-h(\rho_0^\pm)  &{\rm (b)}\\[2mm]
D(\dgv\pm 2\tau\|p) & R>1-h(\rho_0^\pm). &{\rm (c)}
\end{array}\right. 
\end{equation*}
We then have the following result whose proof is given in the
appendix.
%
%
\begin{theorem}\label{thm:tradeoff}
Let $R\ge 1-h(0.5-\tau),$ then the exponent of the undetected error is
bounded below as
\begin{equation}\label{eq:error}
E_e(R,p,\tau)\ge M_+.
\end{equation}
Let $R\ge0, \tau\le p/2,$ then the erasure exponent is bounded below as
\begin{equation}\label{eq:erasure}
E_x(R,p,\tau)\ge M_-.
\end{equation}
\end{theorem}

Remark \ref{remark} applies to this theorem as well: the claim of the theorem
is nontrivial for code rates below $1-h(p+2\tau).$

For $\tau=0$ the bounds also reduce to $E_0(R,p),$ as
expected. However, they are strictly greater that the bounds of Theorem
\ref{thm:bz}. For instance, for the case (c) this can be proved using
 the fact that
$D(\delta\|p)$ is a $\cup$-convex increasing function of $\delta$ for
$\delta>p$:
\begin{align*}
M_+&=D(\dgv(R)+2\tau\|p)>D(\dgv(R)\|p)+2\tau D'_\delta(\dgv(R)\|p)
\remove{
=T(\dgv,p)+2\tau\nu-h(\dgv+2\tau)&\\
&\ge T(\dgv,p)+2\tau\nu-h(\dgv)-2\tau h'(\dgv)
\quad(\text{by concavity of } h)\\
&=D(\dgv\|p)+2\tau(\nu-\log_2(1-\dgv)/\dgv),}
\end{align*}

\begin{figure}[t]
\begin{center}
\setlength{\unitlength}{1mm}
\begin{picture}(88,70)
\medskip\put(25,107){\epsfysize=90mm 
\epsffile[72 400 840 720]{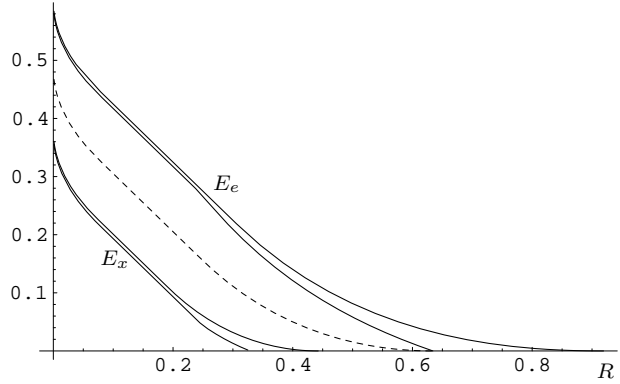}}
\put(84,10){\scriptsize $R$ }
\put(18,25){\scriptsize $E_x$ }
\put(33,35){\scriptsize $E_e$ }
\end{picture}
\begin{minipage}{.8\linewidth}
\caption{Error bounds from Theorems \ref{thm:bz} and \ref{thm:tradeoff}
$(p=0.07, \tau=0.03).$ In each pair the better bound is from 
Thm. \ref{thm:tradeoff}. The dashed line is the function $E_0(R,p)$.}
\label{fig:plot}
\end{minipage}\end{center}
\end{figure}

It is also easy to establish similar inequalities in the other cases.
Typified behavior of the bounds on
$E(R,\rho,\tau)$ from Theorems \ref{thm:bz} and \ref{thm:tradeoff} 
is shown in Fig. \ref{fig:plot}.
These theorems and the other results
in the binary case extend in a standard way to binary-input output-symmetric
discrete memoryless channels and to the $q$-ary symmetric channel, $q\ge 2.$

\begin{remark} The conditions $R\ge 1-h(1/2-\tau)$ and 
$\tau\le p/2$ seem to make Theorem 
\ref{thm:tradeoff} sound more restrictive than Theorem \ref{thm:bz}.
It is possible to remove these conditions and prove somewhat weaker
bounds which will still improve upon Theorem \ref{thm:bz}. However,
the first of the two conditions for small $\tau$ is not a substantial 
restriction of the range of codes rates:
for instance, for $\tau=0.03$ the bound on $E_e$ is valid for
all code rates $R\ge 0.0025.$
Furthermore, it is often the case that the bounds 
(\ref{eq:Ee-low})-(\ref{eq:Ex-high}) are void 
while Theorem \ref{thm:tradeoff} claims nontrivial results. 
For instance, for $(p,\tau)=(0.2,0.09)$ the bound (\ref{eq:Ex-high}) and hence
the rest of Theorem \ref{thm:bz} is trivial while Theorem \ref{thm:tradeoff}
gives nontrivial exponential error bounds for small values of the code rate.
\end{remark}

\begin{remark}
As indicated above, Theorem \ref{thm:bz} can be obtained by a small 
modification of the proof of the our result. 
Generally, Theorem \ref{thm:bz} claims results
weaker than those in (\ref{eq:error})-(\ref{eq:erasure}) because  
the authors of \cite{blo82} in their derivation relied on a suboptimal 
decision region.
\end{remark}

\begin{remark}\label{remark:elias} We can add some details on typical 
error events in the course of decoding. For instance, consider the error-only
case. Let $\rho_{\text{typ}}$ be the relative weight of error vectors that lead to
a decoding error, and let $\omega_{\text{typ}}$ be the relative weight of code
vectors obtained as a result of incorrect decoding. From the proof
it is clear that for the case (a), $\rho_{\text{typ}}=(1-\dgv)p+\dgv/2+\tau,
\omega_{\text{typ}}=\dgv.$ For the case (b), 
$\rho_{\text{typ}}=\rho_0,\omega_{\text{typ}}=\omega_0.$ Finally, for the 
case
(c), $\rho_{\text{typ}}=\dgv, \omega_{\text{typ}}=2\dgv(1-\dgv)+2\tau(1-2\dgv).$
A more detailed discussion of these results for $\tau=0$
is provided by \cite{bar02b}.
\end{remark}

\begin{remark}
Note an alternative expression for the case (b) of $M_\pm$ 
\[
M_\pm=1-R-h(\omega_0)-\omega_0h\Big(\frac12\pm\frac\tau{\omega_0}\Big)-
\frac{\omega_0}2\log_2u\pm\nu\tau.
\]
We stress that the dependence of the bound on $\tau$ is essentially
nonlinear, contrary to the closed-form bounds in \cite{for68b},\cite{blo82}.  
\end{remark}

\section{Related results: The binary case}\label{sect:rH}
1. Let us address the question of error bounds for a specific code
under max-likelihood decoding.
Let $C$ be a code with distance distribution $A_w=2^{n\alpha(\omega)}$ 
$(\omega=w/n, 0\le w\le n)$ and let
\[
K(C)=\max_{1\le w\le n}\frac{A_w}{\max(1,\cA_w)},
\]
$\kappa (C):=n^{-1}\log_2K(C).$ For simplicity only we put $\tau=0.$
The following bound is straightforward.

\begin{theorem} 
\begin{equation}\label{eq:shu-fe}
n^{-1}\log\frac 1{P_{de}(C)}\ge \max(D,E_0(R,p)-\kappa(C))-o(1),
\end{equation}
where
\[
D=-\max_{0<\omega\le 1}\big(\alpha(\omega)+
(\omega/2)\log_2 (4u)\big).
\]
\end{theorem}
\proof
Denote by $P_e(w)$ the error probability
under the condition that the decoded vector is $w$ away from the 
transmitted one. Then
\[P_{de}(C)\le  \sum_{w=1}^n A_w P_e(w)\]
Taking logarithms and switching to exponents, we obtain the first part
of the claim. The second part is equally obvious because
\begin{align*}
P_{de}(C)&\le  \sum_{w=1}^n A_w P_e(w)\le 
2^{-n(1-R)}K(C)\sum_{w=d}^n \binom nw P_e(w)\\
&\le 2^{n(\kappa(C)-E_0(R,p)-o(1))}.
\end{align*} \vskip -8mm\qed

Note that if $R_e\le R<R+\kappa(C)\le R_c,$ then
$E_0(R,p)$ is given by a linear function with slope $-1$, and we can write
$E_0(R,p)-\kappa(C)=E_0(R+\kappa(C),p).$ The second
part of the bound under the maximum in (\ref{eq:shu-fe}) is the main 
result of \cite{shu99}, see Theorem 1 of that paper]. 
The above proof is a shorter way to obtain it.

Note also that for low $R$ bound $D$ on the error rate of $C$ can be better
(and is never worse) than  the second part of (\ref{eq:shu-fe}).
This is due to the fact that in $D$ we maximize the product of
the weight profile and the pairwise error probability, while
in the second bound the maximization of these two terms is separate.

2. Consider the decoding procedure $\tilde\psi_t$ under which
$\tilde\psi_t(\bfx)=\bfc$ if $\dist (\bfc,\bfx)\le t$ and $\tilde\psi_t(\bfx)$
undefined if
such a code vector does not exist. In this case the calculation of 
the error exponent is cumbersome, and depends on the relation between 
$p$ and $t$. One particular case is easy to analyze.
\begin{proposition} Let $\cC$ be a linear binary code with weight
distribution $\cA_i, i=0,\dots,\linebreak[2]n.$ 
Suppose that for every $d\le w\le n$ 
the maximum of 
\[
\sum_{i=\max(\lceil w/2\rceil,w-t)}^n\sum_{\ell=0}^t
\binom wi\binom{n-w}\ell p^{i+\ell}(1-p)^{n-i-\ell}
\]
is attained for $\ell=0, i=w-t.$ Then
$ -n^{-1}\log {P_{de}(C)}\ge \tilde E_e(R,p,\tau)-o(1),$ where
\begin{equation*}\label{eq:bounded-exponent}
\tilde E_e(R,p,\tau)=
\begin{cases}
-\delta h(\tau/\delta)-T(\delta-\tau,p), &0\le R\le1- h(p+\tau(1-p))\\
1-R-h(\tau)-\tau\log_2(1-p)&1- h(p+\tau(1-p))\le R.
\end{cases}
\end{equation*}
\end{proposition}
\proof (outline) We have
\[
P_{de}(C)\le (t+1)^22^{-n(1-R)}\sum_{w=d}^n\binom nw\binom{w}{w-t}
p^{w-t}(1-p)^{n-w+t}.
\]
In the sum on $w$ the summation term is maximized for 
$w\approx n(p+\tau(1-p)).$
The exponent in question is obtained by computing the logarithms and depends
on the sign of $p+\tau(1-p)-\delta.$ For $p+\tau(1-p)\le\delta$
the dominating term is the one with $w=d.$ Upon simplification
we obtain the first case of the claimed bound. Otherwise the maximum
is within the summation range. Taking logarithms, substituting
$\omega=p+\tau(1-p)$ and simplifying, we obtain the second case. 
\qed

In particular, let
$t=0$, which corresponds to the case of pure error detection.
Then  $\tilde E_e(R,p,\tau)$ reduces to the well-known 
lower bound on the exponent of undetected error  \cite{lev77}.

\section{Spherical codes}\label{sect:spherical}

In this section we address the problem of error bounds for erasure
decoding for the case of spherical codes. We assume transmission
over a Gaussian channel with signal-to-noise ratio $A$.
Let $S^{n-1}(r,\bfx)$ be the sphere in $\reals^n$ of radius $r$
with center at $\bfx$. We will write $S^{n-1}(r)$ for $S^{n-1}(r,{\bf 0}).$

Let $X=S^{n-1}(\sqrt {An})$ and 
let $\bfy_1,\bfy_2\in X$ be two vectors. 
One way to measure the distance between them is by the angle
$\angle(\bfy_1,\bfy_2)$, and we will write $\dist(\bfy_1,\bfy_2)=\phi$
if this angle equals $\phi.$ The distance of a code $C\subset X$
is defined in the usual way as the minimum pairwise distance in $C$.
For a given vector $\bfx$ if $\bfy=\bfx+\bfz$ and $\dist(\bfy,\bfx)=\phi,$ 
we will say that $\bfz$ has weight $\phi.$ 

For a code $C\in X$ let $M$ be its size, $R=n^{-1}\ln M$ its rate
and $\theta=\theta(C)$ its distance. We also define the distance
distribution of $C$ as follows:
\[
B(s,t)=M^{-1}\big|\big\{\bfx,\bfx'\in C:\; s\le \dist(\bfx,\bfx')<t\big\}
\big|
\]
and $B(s)=B(0,s),$ so that $M=\int_0^\pi dB(x).$ Given a family of codes,
we call the function $b(x)$  its {\em distance profile} if 
\[
b(x)=\lim_{\genfrac{}{}{0pt}{2}{n\to\infty}{\epsilon\to 0} }
(1/n)\ln B(x-\epsilon,x+\epsilon)
\]
assuming that the limit exists. Throughout this and the next section
we use the notation $\theta_s=\theta_s(R):=\arcsin e^{-R}.$

The decoding mapping for $C$ is defined as follows: $\psi_\tau(\bfy)=\bfx$
if $\dist(\bfy,\bfx')- \dist(\bfy,\bfx)\ge 2\tau$ for all $\bfx'\in C,
\bfx'\ne\bfx$. If such code vector $\bfx$ does not exist, decoding
results in an {\em erasure}. Assume that the transmitted vector
$\bfx$ is displaced by a noise vector $\bfz$ whose coordinates
are i.i.d. Gaussian random variables with mean $0$ and unit variance.
Let $E_e(R,A,\tau)$ and $E_x(R,A,\tau)$ be the best 
attainable exponents of the error and erasure rate, respectively. 
\remove{Formally, $E_e$ is defined as follows: let 
\(
E(C)=-\ln P_{de}(C) 
\)
where $P_{de}(C)$ is the average error probability of $C$ under $\psi_\tau$;
let $E(S^{n-1},M)=\max_{|C|=M} E(C)$ and let 
$E_e(R,A,\tau)=\liminf n^{-1} E(n,e^{Rn})$
with the limit computed over all sequences of codes of asymptotic
rate at least $R$, and $E_x$ is defined along the same lines.}
When $\tau=0$, this is the usual complete decoding, and $E_e$ is the
reliability function of the Gaussian channel. 
In this case we will omit $0$ from our notation and write $\psi, E(R,A).$
The following lower bound on $E(R,A)$ is classical
\cite{sha59}: let $\theta=\theta_s(R),$ then
$E(R,A)\ge E_0(\theta,A),$ where
\begin{equation}\label{eq:gaussian-reliability}
 E_0(\theta,A)=\left\{\begin{array}{l@{\quad}l@{\qquad\qquad}c}
\frac A4(1-\cos\theta), &\frac\pi2\ge \theta\ge \theta_e, &{\rm (a)}\\[2mm]
\frac A4(1-\cos\theta_e)+\ln\frac{\sin\theta}{\sin\theta_e}, 
&\theta_e\ge \theta\ge \theta_c, &{\rm (b)}\\[2mm]
E_{sp}(\theta,A), &\theta_c\ge \theta\ge \arccot\sqrt A, &{\rm (c)}
\end{array}\right.
\end{equation}
where $\csc^2\theta_e=\frac12+\frac12\sqrt{1+\frac{A^2}4},$
$\csc^2\theta_c=\frac12+\frac A4+\frac12\sqrt{1+\frac{A^2}4}$
\begin{eqnarray*}
E_{sp}(\phi,A)&:=&\frac A2-\frac{\sqrt{A}}2 g(\phi)\cos\phi
-\ln(g(\phi)\sin\phi),\\
g(\phi,A)&:=&\frac12(\sqrt A\cos\phi+\sqrt{A\cos^2\phi+4}).
\end{eqnarray*}
This bound will follow as a special case of our derivation.

The (Shannon) volume, or sphere packing
bound \cite{sha59}
establishes the existence of codes of rate $R$ with distance 
arbitrarily close to
$\theta_s(R)$. It is also straightforward to prove that
there exists a code $\cC$ of rate $R$ with distance $\theta_s$ and 
distance distribution
\[
B(s)\le p(n)e^ 
{n(R+\ln\sin\theta)} \qquad(\theta_s\le\theta\le \pi-\theta_s),
\]
where $p(n)$ is some polynomial function. This distribution is induced
by the (normalized) uniform measure on $S^{n-1}$ and therefore plays the
role analogous to that of the binomial distribution in the Hamming space.
The distance profile corresponding to it is $\beta(R,\theta)=R+\sin\theta.$
We will examine the behavior of the error rate with decoding
$\psi_\tau$ being applied to sequences of codes $\cC$ with these properties.

Below we track only one of the two cases, the error-only event,
and state results for the erasure rate sparing the reader the
detailed analysis.
Our goal will be to establish the following theorem. 
%
%
\begin{theorem} \label{thm:spherical} 
Let $R$ be the code rate, let $\theta_s=\theta_s(R)$ be the code distance
and let $\tau>0.$
The exponent $E_e(R,A,\tau)$ 
is bounded below by $M(R),$ where
for $\pi/2\ge\theta_s>\theta_1$
\begin{equation}\label{eq:sphere-expurgation}
M(R)=\frac A4(1-\cos(\theta_s+\tau))- G(\theta_s,\tau),
\end{equation}
for $\theta_1\ge\theta_s>\theta_2 $
\begin{equation}\label{eq:sphere-straight}
M(R)=\frac A4(1-\cos(\theta_1+\tau))+\ln\frac{\sin\theta_s(R)}
{\sin\theta_1}- G(\theta_1,\tau),
\end{equation}
and for $\theta_2> \theta_s$
\begin{equation}\label{eq:sphere-spherepacking}
M(R)=E_{sp}(\rho,A).
\end{equation}
Here 
\begin{equation}\label{eq:G}
G(\phi,\tau)=\frac12\ln\Big[1+\frac{A\cos^2\frac{\phi+\tau}2
(\sin^2\frac{\phi+\tau}2-\sin^2(\frac\phi2+\tau))}
{\cos^2(\frac\phi2+\tau)}\Big]
\end{equation}
$\theta(x)$ is the (real, positive-valued) function defined 
implicitly by the equation
\begin{equation}\label{eq:Elias-angle}
\cot\theta=\frac{\cos^2 x\tan(\frac\theta2+\tau)}
{\cos(\theta+2\tau)-\cos 2x} \quad(0\le x\le\pi/2),
\end{equation}
$\rho=\rho(R) \in [t_s,2t_s]$ is the unique angle such that
\begin{equation}\label{eq:decoding-angle}
R+\ln\sin\theta(\rho)+\frac12\ln
\Big(1-\frac{\tan^2((\theta(\rho)/2)+\tau)}
{\tan^2\rho}\Big)=0,
\end{equation}
$\theta_1$ is the root of
\begin{equation}\label{eq:expurgation-angle}
\frac d{dx}\big(\ln\sin x+\frac A4\cos(x+\tau)+G(x,\tau)\big)=0
\end{equation}
and $\theta_2=\theta_s(R^\ast),$ where $R^\ast$ is the root of
$\theta(\rho(R))=\theta_1.$

A lower bound on the exponent $E_x(R,A,\tau)$ is obtained on replacing
$\tau$ by $-\tau$ throughout.
\end{theorem}

\begin{figure}[t]
\begin{center}
\setlength{\unitlength}{1mm}
\begin{picture}(88,70)
\medskip\put(25,107){\epsfysize=90mm 
\epsffile[72 400 840 720]{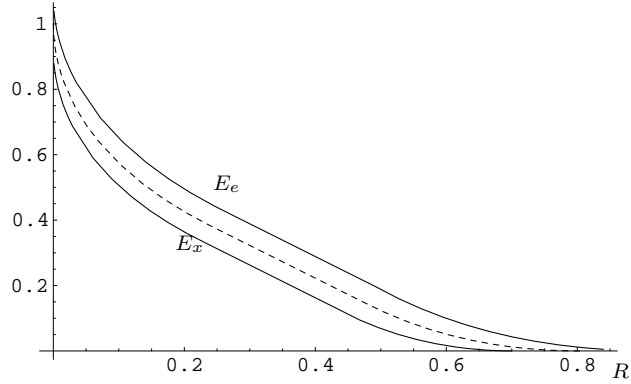}}
\put(86,10){\scriptsize $R$ }
\put(28,27){\scriptsize $E_x$ }
\put(33,35){\scriptsize $E_e$ }
\end{picture}
\begin{minipage}{.8\linewidth}
\caption{Error bounds from Theorem \ref{thm:spherical} $(A=4,\tau=0.04\approx
2.3^\circ)$.
The dashed line is the function $E_0(R,A)$.}
\label{fig:plot-sp}
\end{minipage}\end{center}
\end{figure}

Although it is not immediately seen, for $\tau=0$ we have $M(R)=E_0(R),$
so in this case the bounds simplify significantly. 
For instance, $G(\phi,0)=0,$ and the bound (\ref{eq:sphere-expurgation}) 
reduces to (\ref{eq:gaussian-reliability}a), 
the value $\rho$ equals $\theta_s(R)$, the angle $\theta_1$ is simply 
$\theta_e$ of (\ref{eq:gaussian-reliability}a-b), and so on. 
We explain these and indicate further connections with bound 
(\ref{eq:gaussian-reliability}) in remark \ref{remark:sphere-elias}
below.  
Note that though there seems to be no closed-form expression for the
exponents, it is easy to compute them for any given $A,\tau.$ 
It helps to observe that on substituting $\theta(\rho)$ into 
(\ref{eq:decoding-angle}), this equation contains only one unknown, $\rho.$
We show the behavior of the bounds in Fig. \ref{fig:plot-sp}. Note that
$M(\rho)> 0$ for $0\le\rho<\arccot\sqrt A.$
Note also that $G(\phi,\tau)$ is negative (and usually small),
so on omitting it from expressions (\ref{eq:sphere-expurgation}),
(\ref{eq:sphere-straight}) we still obtain valid lower bounds.

The remaining part of this section is devoted to the proof of this
theorem.
We begin with some notation and technical results.
Let $\Con (\bfx,\phi)$ denote the circular cone with apex at the
origin, axis given by a vector $\bfx\in \reals^n$ and solid half-angle $\phi.$
We write $f(n)\logasymp g(n)$ if $\lim_{n\to\infty}
\frac1n\ln\frac{f(n)}{g(n)}=0.$

We will need the following lemmas.
\begin{lemma}\label{lemma:Q} {\rm\cite{sha59}}
Let $\bfx\in X$ and $\bfz$ a random Gaussian vector.
Let $Q(\phi)$ be the probability that $\bfx+\bfz\not\in \Con(\bfx,\phi).$
Then $Q(\phi)\logasymp e^{-n E_{sp}(\phi,A)}$ and $-dQ(\phi)\logasymp
e^{-n E_{sp}(\phi,A)}d\phi.$
\end{lemma}

\begin{lemma}\label{lemma:cap} {\rm \cite{sha59}}
Let $T(\phi)$ be the area of the spherical
cap on the sphere $S^{n-1}(r)$ cut out by a cone $\Con(\bfx,\phi)$. Then 
as $n\to\infty$
\[
T(\phi)\sim \frac{2\pi^{(n-1)/2} r^{n-1}\sin^{n-1}\phi}
{(n-1)\Gamma\big(\frac{n-1}2 \big)\cos\phi}.
\]
For the normalized area $\Omega(\phi)=T(\phi)/T(\pi)$ we have
\[
\Omega(\phi)\logasymp (\sin\phi)^n.
\]
\end{lemma}

\begin{lemma} \label{lemma:Laplace-method} {\rm (e.g. \cite[p.65]{bru58})}
{\rm(Laplace method)} 
Let 
\[
h(\lambda)=\int_a^b e^{\lambda q(x)}dx  \quad(-\infty\le a<0<b\le\infty).
\]
Suppose that the integral converges absolutely at least for 
sufficiently large $\lambda$.

{\rm (i)} Suppose that 
the absolute maximum of $q(x)$ in $[a,b]$ is attained for $x=0$, 
that $q'(0)$ exists and is continuous in some neighborhood of $0$,
and that $q''(0)<0.$ Then
\[
h(\lambda)\sim \sqrt{-\frac{2\pi}{\lambda q''(0)}}e^{\lambda q(0)}
\quad(\lambda\to\infty).
\]

{\rm (ii)} Suppose that $a=0$ and that the absolute maximum of $q(x)$ 
in $[a,b]$
is attained for $x=0.$ Then 
\[
h(\lambda)\sim -
\frac{e^{\lambda q(0)}}{\lambda q'(0) } 
\quad(\lambda\to\infty),
\]
provided that $q'(0)<0.$ 
\end{lemma}

Shannon's approach to bounding the rate of error events is as follows.
Let $\cE$ denote one of the two events: error, or error or erasure.
\begin{lemma}\label{lemma:shannon}
{\rm \cite{sha59}} Let $\bfz$ be the channel error vector.
Then
\[
P(\cE,C)\le \min_{\rho} P(\cE|w(\bfz)\le\rho)+P(w(\bfz)\ge\rho).
\]
\end{lemma}

Generally, the minimum is attained for different $\rho$ depending on the
meaning of $\cE.$ Note that to obtain a valid bound we do not have
to optimize on $\rho,$ taking an arbitrary value at our convenience.
Below we always assume that $\rho<90^\circ.$

\begin{figure}[t]
\begin{center}
\setlength{\unitlength}{1mm}
\begin{picture}(88,70)
\medskip\put(0,97){\epsfysize=70mm 
\epsffile[72 400 840 720]{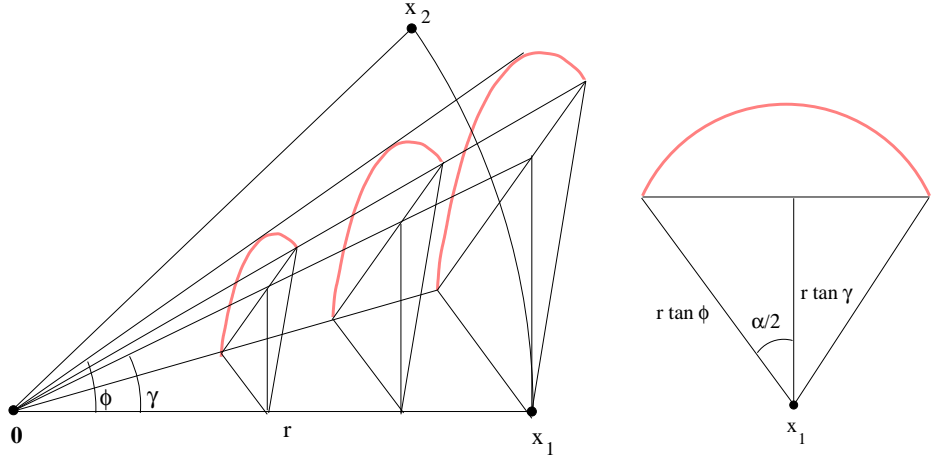}}
\end{picture}\vskip-1cm
\begin{minipage}{.8\linewidth}
\caption {Derivation of Lemma \ref{lemma:cones} }
\label{fig:cones}
\end{minipage}\end{center}
\end{figure}

\begin{lemma} \label{lemma:cones}
Let $\bfx_1\in C$ be transmitted and let 
$P_\theta(\bfx_1\to\bfx_2)$ be the probability
that decoding $\psi_\tau$ mistakes $\bfx_1$ for a fixed code vector 
$\bfx_2$ with $d(\bfx_1,\bfx_2)=\theta$. Then 
$P_\theta(\bfx_1\to\bfx_2)\logasymp F(\theta,\tau),$ where
\[
F(\theta,\tau):= \int\limits_{\theta/2+\tau}^\rho 
\Big(1-\frac{\tan^2(\theta/2+\tau)}{\tan^2\phi}\Big)^{n/2}
e^{-n E_{sp}(\phi,A)}d\phi.
\]
\end{lemma}
\proof Let $\bfz$ be the error vector with $w(\bfz)=\phi, \|\bfz\|=r.$ 
Let us compute the fraction of such errors that lead to a decoding
error that outputs $\bfx_2.$ For this to happen it suffices that
$\phi\ge \theta/2+\tau$ and $\|\bfx_2-\bfy\|\le\|\bfx_1-\bfy\|,$ where
$\bfy=\bfx_1+\bfz$ is the received vector. This fraction equals
the normalized area of the spherical cap cut out on the surface
of $\Con(\bfx,\phi)$ by the hyperplane perpendicular to $\bfx_1$ and
located at a distance $r$ from the origin. Taking in Fig. \ref{fig:cones}
$\gamma=\theta/2+\tau$, we compute for the angle $\alpha$ of this cap
\[
\sin^2\alpha/2=1-\big(\frac{r\tan(\theta/2+\tau)} {r\tan\phi}\big)^2=
1-\frac {\tan^2(\theta/2+\tau)}{\tan^2\phi}
\]

The normalized area $\Omega$
of the cap in question is given by Lemma \ref{lemma:cap}:
\[
\Omega\logasymp (\sin\alpha/2)^n
\]
and does not depend on the distance $r$ from the origin. Hence we
may integrate $r$ out and obtain for the differential probability
\[
P(\bfx_1\to\bfx_2|\phi\le w(\bfz)\le\phi+d\phi)\logasymp
-\big(1-\frac {\tan^2(\theta/2+\tau)}{\tan^2\phi}\big)^{n/2} dQ(\phi)
\]
Now the claim of the lemma is obtained by integrating on $\phi$
from $\theta/2+\tau$ (because errors of smaller weight cannot
lead to decoding error) to $\rho$ (because errors of greater
weight are assumed to always lead to a decoding error)
and substituting $dQ(\phi)$ from Lemma \ref{lemma:Q}.
\qed

Putting the pieces together, we obtain the following bound
on the probability of error
for spherical codes under $\psi_\tau$. 
\begin{theorem}\label{thm:error-bound}
Let $C$ be a code of rate $R$ and distance $\theta$
with distance profile $b(\theta)$.
Then for any $\rho\in(\theta/2,\pi/2)$
\begin{equation}\label{eq:main-bound}
P(\cE;C)\lesssim \int_{\theta}^{2(\rho-\tau)} e^{nb(\theta)}F(\theta,\tau)d\theta
+Q(\rho).
\end{equation}
\end{theorem}

\proof Follows on applying Lemmas \ref{lemma:shannon} and
\ref{lemma:cones} and the union bound.\qed

Now we are ready to complete the proof of Theorem \ref{thm:spherical}.
First let us find the asymptotic behavior of $F(\theta,\tau)$.
We have $F(\theta,\tau)\logasymp\int e^{nq(\phi)}d\phi$, where 
\[
q(\phi)=\frac12\ln\Big(1-\frac{\tan^2(\theta/2+\tau)}{\tan^2\phi}\Big)
- E_{sp}(\phi,A).
\]
From the equation $q'(\phi)=0$ we find that
the maximum of the integrand is attained for $\phi_0(\tau)$ defined
by
\begin{equation}\label{eq:phi0}
\sin^2\phi_0=\frac{4+A\sin^2(\theta+2\tau)}{2(2+A+A\cos(\theta+2\tau))}.
\end{equation}
The asymptotic value of the integral is obtained by Lemma 
\ref{lemma:Laplace-method}
and depends on the location of $\phi_0$ with respect to the integration
limits. First, it is easy to see that $\phi_0>\theta/2+\tau$ for
any $0<A<\infty, 0<\theta<\pi-2\tau.$ Indeed, it suffices to show that
$\sin^2\phi_0>\sin^2(\theta/2+\tau).$ Therefore, compute
\begin{align*}
2(2+A+&A\cos(\theta+2\tau))(\sin^2\phi_0-\sin^2(\frac\theta2+\tau))\\
&=4+A\sin^2(\theta+2\tau)-2\sin^2(\frac\theta2+\tau)
(2+A+A\cos(\theta+2\tau))\\
&=4(1-\sin^2(\frac\theta2+\tau))>0.
\end{align*}
It remains to examine the location of $\phi_0$ with respect to the 
upper limit of integration, $\rho$.
We have the two following cases.

1. $\phi_0<\rho.$ Then by Lemma \ref{lemma:Laplace-method}(i) the 
behavior of the integral is determined by $\phi$ in the
neighborhood of $\phi_0.$ We obtain
\[
n^{-1}\ln F(\theta,\tau)\sim \frac12\ln \Big(1-\frac{\tan^2(\theta/2+\tau)}
{\tan^2\phi_0}\Big)-E_{sp}(\phi_0,A).
\]
Next let us proceed to computing the asymptotic expression for the
outer integral in (\ref{eq:main-bound}).
Denoting the integrand by $D$, 
substituting the value of $\phi_0$ and taking $b(\theta)=\beta(R,\theta)$, 
after all simplifications, we arrive at the expression
\[
-n^{-1}\ln D\sim \frac A4(1-\cos(\theta+\tau))-\beta(R,\theta)-G(\theta,\tau).
\]
Now invoke again Lemma \ref{lemma:Laplace-method}. The main term
of the integral depends on the relative location of the 
maximizing value of $\theta,$ denoted by $\theta_1,$ and the
integration limits. 
As it turns out, for $\theta_1$ we have $0<\theta_1<2\rho-2\tau ,$ so
what matters is the mutual location of $\theta_1$ and $\theta_s.$ If 
$\theta_1<\theta_s,$ then by Lemma \ref{lemma:Laplace-method}(ii)
the main term is determined by $\theta=\theta_s.$ Since $\beta(R,\theta_s)=0$,
we obtain for $E_e(R,A,\tau)$ the bound 
\[
E_e(R,A,\tau)\ge\frac  A4(1-\cos(\theta_s+\tau))-G(\theta_s,\tau),
\]
which is (\ref{eq:sphere-expurgation}). 
On the other hand, if $\theta_1\ge \theta_s,$ then
we use part (i) of the same lemma and obtain the bound 
(\ref{eq:sphere-straight}).
This proves the first two parts of the theorem except the upper limit 
$\theta_2$ of range of angles in (\ref{eq:sphere-straight}) 
which will be established later.

2. $\phi_0\ge\rho.$ Now the asymptotic value of $F(\theta,\tau)$ is 
determined by $\phi=\rho,$  and so
\[
n^{-1}\ln F(\theta,\tau)\sim \frac12\ln \Big(1-\frac{\tan^2(\theta/2+\tau)}
{\tan^2\rho}\Big)-E_{sp}(\rho,A).
\]
We proceed to computing the asymptotic expression for the
outer integral in (\ref{eq:main-bound}). Again taking 
$b(\theta)=\beta(R,\theta)$ and denoting the integrand by $D,$ we obtain
\[
n^{-1}\ln D\sim R+\ln\sin\theta+\frac12\ln \Big(1-
\frac{\tan^2(\theta/2+\tau)}{\tan^2\rho}\Big)-E_{sp}(\rho,A).
\]
Differentiating, we find that the maximum of this expression on 
$\theta$ is attained for $\theta=\theta(\rho),$ and it is possible
to prove that with our choice of $\rho$ the value $\theta(\rho)$
is always within the integration range: 
$\theta_s\le \theta(\rho)<2(\rho-\tau).$
Furthermore, by (\ref{eq:decoding-angle}) the first three terms in the 
expression for $D$ add up to zero.
Concluding, in this case the integral on $\theta$ evaluates 
asymptotically to
\[
\int_{\theta_s}^{2\rho-2\tau} e^{n(\beta(R,\theta)+\ln F(\theta,\tau))}d\theta
\logasymp e^{-n E_{sp}(\rho,A)}.
\]
Clearly, the second term in (\ref{eq:main-bound}) has the same asymptotic
behavior, which is therefore the answer in the case studied.
It remains to find the value $\theta_2$ when the main term of the
estimate moves from (\ref{eq:sphere-straight}) to 
(\ref{eq:sphere-spherepacking}). This obviously happens when the
two functions first become equal  as the angle $\theta_s$ decreases
from $\theta_1$ or when the rate $R$ reaches 
the value such that $\theta(\rho(R))=\theta_1,$ or when $\theta_s=\theta_2.$
This concludes the proof of (\ref{eq:sphere-spherepacking}) and
thus of the theorem. \qed

\begin{remark}\label{remark:sphere-elias} The results of \cite{sha59}
are obtained from this theorem by substituting $\tau=0$ in 
(\ref{eq:Elias-angle}). Denoting 
$\theta(\rho)$ in
this case by $\theta_E$, we find that $\cos\theta_E=\cos^2\rho.$
Further, substituting $\theta_E$ into (\ref{eq:decoding-angle}), we find
that $\rho(R)=\theta_s(R),$ i.e., the optimizing value of the decoding
radius in (\ref{lemma:shannon}) in this case is $\theta_s$.
Taking $\tau=0$ in (\ref{eq:expurgation-angle}), we obtain for
$\theta_1$ the explicit equation $\cos\theta_1=(A/4)\sin^2\theta_1$
whence $\theta_1=\theta_e.$ Further, the equation for $R^\ast$ which in
general is $\theta(\rho(R))=\theta_1,$ now reduces to $\theta_E=\theta_1$ or
   
\begin{equation}\label{eq:te}
\csc^2\theta_E=[\sin^2\theta_s(R^\ast)(2-\sin^2\theta_s(R^\ast))]^{-1}=
\frac12\Big(1+\sqrt{1+\frac{A^2}4}\Big).
\end{equation}
From this we find
$R^\ast=-\ln\sin\theta_c,$ or $\theta_2=\theta_c$ of 
(\ref{eq:gaussian-reliability}b-c). Hence $\theta_2$ equals 
the critical angle and $R^\ast$ equals the critical rate of the channel.

These remarks also enable us to make some observations on typical
error events in the course of decoding of codes $\cC$. 
They are easier understood for $\tau=0.$ Since
the codes $\cC$ generally are not distance invariant, the following
is valid on average only.
\ol \item Suppose that $\pi/2>\theta_s>\theta_e,$ then the 
errors that contribute to the main term of $E(R,A)$ most probably
are of weight $\phi_0\in(\theta_s/2,\theta_s).$ In the case of
decoding error the
typical distance of the output code vector from the transmitted one 
equals $\theta_s$ and does not depend on the level of noise in the channel.

\item For $\theta_e>\theta_s>\theta_c,$ typical errors are also of
weight $\phi_0$ and result into code vectors at distance $\theta_e$
from the transmitted one. From the moment that $\theta_E=\theta_e$
or $\theta_c=\theta_s,$ typical errors are of weight $\theta_s$ and
the resulting code vectors are distance $\theta_E$ from the transmitted one.
The error rate in this case does not depend on the actual channel noise or 
on the distance of the code.
\eol
\end{remark}

\begin{remark} (The Elias angle). Note that the value $\theta_E$
gives the answer to the following geometric question. Let 
$\bfx\in \cC$ be a vector in a code of rate $R$ and distance $\theta_s.$
Consider all the neighbors $\bfx'$ of $\bfx$ in $\cC$ such that 
$\dist(\bfx,\bfx')=\alpha$ for a given $\alpha$ and draw the cones
$\Con(\bfx',\theta_s)$ about them. What is the minimum value of 
$\alpha$ such that fraction of the surface of $\Con(\bfx,\theta_s)$ 
covered by these cones asymptotically becomes one? The answer follows from
(\ref{eq:decoding-angle}) and is given by $\alpha=\theta_E$.
This parameter plays the same role for $S^{n-1}$ as the Elias radius
for the Hamming space (see, e.g., \cite{bar02b}),
 therefore we call it the {\em Elias angle}. 
This also hints, by the same geometric argument as in the Hamming case
at the bound $\theta\le \theta_E$ for the maximal attainable minimum distance
of a spherical code of rate $R$. 
Solving the first inequality in (\ref{eq:te}) for $R$,
we obtain a different form of this bound, namely
$R\le -\ln (\sqrt2\sin(\theta/2))$. This is an old 
bound of Rankin \cite{ran55} and Coxeter \cite{cox63} on the rate of 
a spherical code of distance $\theta$ whose proof we therefore obtain.
\end{remark}

\begin{remark} Note that though we emphasized codes $\cC$ in our
derivation, many parts of it, such as bound (\ref{eq:main-bound}),
apply to any sequence of codes with a known distance profile.
They are also applicable to binary codes used over the binary-input
Gaussian channel. Let $C$ be a binary spherical code, i.e. a subset of
$S^{n-1}(\sqrt{An})$ such that coordinates of every vector in $C$
take values $\pm\sqrt A$. Let $d(C)$ and $\theta(C)$ be the minimum
Hamming and angular distance in $C$ respectively,
then $d(C)=n(1-\cos\theta(C))/2.$
We can specialize bound (\ref{eq:main-bound}) to this case as follows:
\begin{equation}\label{eq:poltyrev}
P(\cE,C)\le\sum_{w=d(C)}^{\lfloor n(1-\cos2\rho)/2\rfloor} A_w
F(\theta_w,\tau)+Q(\rho),
\end{equation}
where $(A_d,\dots,A_n)$ is the distribution vector of Hamming distances
in $C$ and
$\theta_w=\arccos (1-2w/n).$ It is straightforward to compute the trade-off
bounds analogous to Theorem \ref{thm:spherical}.
For $\tau=0$ they reduce to
a bound on the error rate of complete decoding for $C$ which can
be used for finite length as well. For that purpose,
more accurate approximations on $F(\theta,\tau)$ than those used above
are readily available. In particular, the normalized area of the spherical
cap can be computed with arbitrary precision from the asymptotic
series provided by the Laplace method \cite{olv97}, and a more precise 
expression for $Q(\phi)$ than the one quoted in Lemma \ref{lemma:Q} 
is given in \cite[Eq. (51)]{sha59}. Asymptotically 
(\ref{eq:poltyrev}) becomes the same as Poltyrev's ``tangential-sphere''
bound \cite{pol94a}; for binary linear codes with binomial weight spectrum 
$\cA_i$ (see Sect. \ref{sect:Hamming}) we immediately recover the 
part of the random coding exponent below the cutoff rate.
\end{remark}

\section{Related results: Bounded distance decoding and error detection}

Let us address a related question, that of error exponents for bounded
distance decoding of spherical codes. Consider the following partial
decoding mapping $\tilde\psi_\tau: X\to C:$ if $\bfy$ is within
distance $\tau$ of a code vector $\bfx$, then $\tilde\psi_\tau(\bfy)=\bfx$,
and if there is not such $\bfx$, the value of $\tilde\psi_\tau(\bfy)$
is undefined. Recall that by distance $\dist(\bfx,\bfy)$ we mean the angle
$\angle(\bfx,\bfy)$. Again we are interested in the best attainable 
error exponent of such decoding for spherical codes. 
The following proposition is obvious if in  Lemma \ref{lemma:shannon}
we take $\rho=\pi/2$, and use Lemmas \ref{lemma:cones} and \ref{lemma:Q}.
\begin{proposition} Let $C$ be a code
in $X$ with distance $\theta(C)>0$ and distance profile $b(\theta).$
Then for any $\epsilon>0$ the probability of decoding error
\begin{equation}\label{eq:bounded}
P_{de}(C)\lesssim
\iint
e^{n\big(b(\theta)+
\frac12\ln\big(1-\frac{\tan^2(\theta-\tau)}{\tan^2\phi}\big)\big)}
dQ(\phi)d\theta +e^{-n E_{sp}(\pi/2-\tau-\epsilon,A)},
\end{equation}
where $\theta(C)\le\theta\le\pi/2-\tau-\epsilon$ and
$\max (\theta(C)/2,\theta-\tau)\le\phi\le\theta+\tau.$
\end{proposition}

Note that the choice of the upper limit $\theta<\pi/2-\tau-\epsilon$ 
is forced by Lemma \ref{lemma:cones}.

Let us first study the asymptotic behavior of the integral on $\phi.$
Letting $\tilde F(\theta,\tau)=\int e^{n q(\phi)}d\phi$ with
\[
q(\phi)=\frac12\ln\Big(1-\frac{\tan^2(\theta-\tau)}{\tan^2\phi}\Big)-
E_{sp}(\phi,A)
\]
we find the root $\phi_0$ of $q'(\phi)=0$ to satisfy
\[
\sin^2\phi_0=\frac{4+A\sin^2(2(\theta-\tau))}{4+2A+2A\cos(2(\theta-\tau))}
\]
(cf. (\ref{eq:phi0})). It is easy to see that $\phi_0<\pi/2.$
By the calculation following (\ref{eq:phi0})
we know that also in the present situation $\phi_0>\theta-\tau,$
so the asymptotics of $\tilde F(\theta,\tau)$ depends on the
mutual location of $\phi_0$ and $\theta+\tau.$ 
Thus we obtain for the error exponent
$\tilde E_e(R,A,\tau)=n^{-1}\ln P_{de}(C)$ 
\[
\tilde E_e(R,A,\tau)\gtrsim \max_{\theta(C)\le \theta\le \pi/2-\tau
} (-b(\theta)-q(\theta_0))
\]
where $\theta_0=\phi_0$ if $\phi_0<\theta+\tau$ or $\theta_0=\theta+\tau$
otherwise. The first situation usually occurs for high code rates,
and the last for low rates. As above, the second term in (\ref{eq:bounded})
can improve the high-rate case.

\ignore{ 
The best lower bound over codes of nonzero rate is obtained
by taking $C=\cC$, and so $b(\theta)=\beta(R,\theta).$ However even in this
case the result does not simplify to a closed-form expression.}

We conclude this section with studying 
{\em error detection} with spherical codes.
Generally error detection proceeds as follows: if the received
vector $\bfy$ is contained in $C$, the decoder outputs $\bfy,$ otherwise its
output is undefined. Clearly for any finite-size code $C\subset S^{n-1}$ 
the probability of undetected error is zero, therefore we define 
error detection
as a limiting case of bounded distance decoding and study the behavior
of $P$ as $\tau\to 0.$ Since the code is a finite set, the cumulative
measure of spherical caps about code vectors tends to zero if so does
their angle. Hence the error probability $P_{de}(C)$ is determined
by the decrease rate of the area of a spherical cap. 
Assume that $C$ is a code with distance $\theta$
separated from $0$ and distance profile $b(\theta).$ If $\tau=0,$ 
then by (\ref{eq:phi0}) we have 
\[
\sin^2\phi_0-\sin^2\theta=4(1-\sin^2 \theta)>0;
\]
so by continuity for small positive $\tau$ also 
$\sin^2\phi_0>\sin^2(\theta+\tau).$ Hence we obtain
\[
n^{-1}\ln\tilde F(\theta,\tau)\sim \frac12\ln\Big(1-\frac{\tan^2(\theta-\tau)}
{\tan^2(\theta+\tau)}\Big)-E_{sp}(\theta+\tau,A).
\]
Since for $\tau\to 0$
\[
\frac{\tan^2(\theta-\tau)}{\tan^2(\theta+\tau)}=1-\frac 8{\sin 2\theta}\tau
+O(\tau^2)
\]
we conclude that the probability of undetected error essentially
does not depend on the distance profile of $C$ and behaves as
\[
P_{ue}\logasymp\exp(n\ln\sqrt{8\tau\csc2\theta(C))})=
(8\tau\csc2\theta(C))^{n/2}.
\]
We see that basically one and the same behavior can be claimed for any code 
with minimum distance $\theta$ separated from $0;$ 
thus the asymptotic answer for the 
undetected error rate of spherical codes is known exactly 
(unlike the more difficult
Hamming case where it essentially depends on optimal codes).

{\bf Acknowledgment.} Thanks to an anonymous referee for pointing out a 
potential error in the original derivation.

%
%

\bigskip
{\Large\bf Appendix}

{\sc A proof of Theorem \ref{thm:tradeoff}.}
Let $\cC$ be a binary linear code of rate $R$, distance
$d$ and weight distribution $A_i(\cC)$, where
$A_i(\cC)\logasymp \cA_i$. The weight profile of $\cC$
has the form $\alpha_0(\omega):=h(\omega)-h(\dgv).$
Let $F_+$ denote the undetected error event and $F_-$ the
error-or-erasure event. Assume w.l.o.g. that the transmitted vector is all-zero
and that $\bfe$ is the channel error vector. The probability of the
error events can be bounded above as follows:
\[
P(F_\pm)\le P(F_\pm|\bfe\in S_{r^\pm}(0))+P(\bfe\not\in S_{r^\pm}(0))
\]
for some positive $r^+$ and $r^-.$ Below we choose $r^\pm=d\pm 2t.$
More concretely, we have
\begin{equation}\label{eq:Sr0}
P(F_\pm|\bfe\in S_{r^\pm}(0))\le \sum_{w=d}^{2r^\pm \mp 2t} A_w(\cC)
\sum_{e=w/2
\pm t}^{r^\pm}
p^e(1-p)^{n-e}\sum_{s=0}^{e\mp 2t} p_{e,s}^w
\end{equation}
More accurately, the range of the summation index $e$ in the above expression
is $w\ge \lceil w/2\rceil \pm t$ if $w$ is odd and $w/2+1\pm t$
if $w$ is even; we will ignore this. Let us proceed with the
undetected error case and rewrite the estimate in an explicit form,
substituting the value of $A_w$:
\begin{align}
P(F_+)\le 2^{-n(1-R)}\sum_{w=d}^{2d+ 2t}\binom nw \sum_{e=w/2+t}^{d+2t}
&p^e(1-p)^{n-e}\sum_{i=\lceil\frac w2\rceil+t}^e \binom wi \binom{n-w}{e-i}
\nonumber\\
&+\sum_{e=d+2t+1}^n\binom nep^e(1-p)^{n-e}.
\label{eq:error-probability-tradeoff}
\end{align}
To facilitate transition to this expression from (\ref{eq:Sr0}) 
notice that if $\bfc$ is 
the incorrect codeword of weight $w>0$ output by the decoder
and $\bfe$ is the error vector then
the index $i=|\supp(\bfe)\cap\supp(\bfc)|.$

The product $\binom wi\binom{n-w}{e-i}$ is maximized for 
\[
i\approx \frac {ew}{n}\le \frac {(d+2t)w}{n}\le \frac w2+ t,
\]
where the last step follows (for large $n$) by the assumption of the
theorem 
$R\ge 1-h(1/2-\tau)$ which translates into $\dgv(R)+\tau\le 1/2.$
Therefore the sum on $w$ in (\ref{eq:error-probability-tradeoff})
for large $n$ can be estimated from above by
\begin{align}
&n\sum_{e=d/2+t}^{d+2t}p^e(1-p)^{n-e} \sum_{w=d}^{2(e-t)}
\binom nw \binom{w}{w/2+t}\binom{n-w}{e-w/2-t}\label{eq:sum-te0}\\
&\cong\sum_{e}\binom ne p^e(1-p)^{n-e}\sum_w \binom{e}{w/2+t}\binom{n-e}{w/2-t}
\label{eq:sum-te}
\end{align}
(since $\binom nw p_{e-2t,e}^w=\binom ne p_{e-2t,w}^e$). In the sum
on $w$ we are counting the number of vectors of weight $w$ which are
distance $e-2t$ away from a given vector of weight $e$. This number
is maximized when 
\[
\frac{e-t-\frac w2}{e}\approx \frac{e-2t}{n}.
\]
Introducing the notation $w=\omega n, e=\rho n,$ we can rewrite this
relation as
\[
\omega^\ast=2\rho(1-\rho)-2\tau(1-2\rho).
\]
Thus the expression in (\ref{eq:sum-te}) is $\cong$-equivalent to
\begin{align}
\sum_e \binom ne &p^e(1-p)^{n-e} \binom{e}{\rho(n-e)+2t\rho}\binom{n-e}
{(\rho-2\tau)(n-e)}\nonumber\\
&\cong \sum_e\binom ne\binom n{e-2t} p^e(1-p)^{n-e}\nonumber \\
&\cong \max_{\dgv/2+\tau\le\rho\le\dgv+2\tau}
\exp[-n (D(\rho\|p)-h(\rho-2\tau))].\label{eq:rce}
\end{align}
The last exponent is maximized for $\rho=\rho_0^+,$ and thus the
unrestricted maximum on $\omega$ is attained for $\omega=\omega_0.$
The cases (a)-(c) of the theorem are realized depending on how
these values are located with respect to the optimization limits
\[
\omega\ge \delta, \quad \dgv/2+\tau\le\rho\le\dgv+2\tau.
\]
If both $\rho_0$ and $\omega_0$ satisfy these inequalities, we substitute
them into (\ref{eq:rce}), recall the factor $2^{Rn-n}$ from
(\ref{eq:error-probability-tradeoff}) and arrive at case (b) of 
the bound $M_+$ in (\ref{eq:error}).

If $\rho_0^+>\dgv+2\tau$ then we substitute $\rho=\dgv+2\tau,\omega=
\omega^\ast$ and obtain the expression
\[
D(\dgv+2\tau\|p)-h(\dgv)+(1-R)=D(\dgv+2\tau\|p),
\]
i.e., case (c). Finally if $\omega_0\le \delta,$ we substitute $w=d$ in 
(\ref{eq:sum-te0}) and obtain
\begin{align*}
2^{Rn-n}&\sum_{e=d/2+t}^{d+2t}p^e(1-p)^{n-e}\binom nd\binom{d}{d/2+t}
\binom{n-d}{e-d/2-t}\\
&\cong \binom{d}{d/2+t}\max_{e\ge d/2+t} 
\binom{n-d}{e-d/2-t}p^e(1-p)^{n-e} \qquad\{a:=e-t\}\\
&=2^{-\nu t}\binom{d}{d/2+t}\max_{a\ge d/2}\binom{n-d}{a-d/2}p^a(1-p)^{n-a}.
\end{align*}
The last maximum is attained for $a-d/2\approx (n-d)p.$ Substituting and
switching to exponents, we arrive at the case (a) in (\ref{eq:error}).

A proof is needed to show that in this case the first of the two terms
in (\ref{eq:error-probability-tradeoff}) provides the dominating
exponent; this is a straightforward calculation which we shall omit.
This completes the analysis of the undetected error event $F_+$.

Let us sketch the proof in the error-and-erasure case $F_-$. 
Now the sum (\ref{eq:Sr0})
can be written as
\[
2^{Rn-n}\sum_{w=d}^{2d-2t}\binom nw\sum_{e=w/2-t}^{d-2t}p^e(1-p)^{n-e}
\sum_{i=w/2-t}^e \binom wi\binom{n-w}{e-i}.
\]
We would like to prove that the maximum on $i$ which is again attained
for $i\approx ew/n,$ at least for large $n$ falls below $w/2-t$.
This will follow from the inequality
  $$
    \omega \delta -\frac \omega 2 \le (2\omega -1)\tau 
  $$
which is proved as follows. We can assume that $\omega <1/2.$
By assumption, $t\le pn/2$ and hence
$t\le d/2$ since $\dgv(R)\ge p$ for $R\le \cC.$ 
Then
  \begin{align*}
    \omega \delta -\frac \omega 2 +(1-2\omega)\tau
        \le \frac12(\delta-\omega)\le 0.
  \end{align*}  
Hence for any $\rho\le\delta-2\tau$ we have
\[
\omega\rho\le\omega(\delta-2\tau)\le\frac\omega 2-\tau
\]
as desired. So instead of (\ref{eq:sum-te0}) we obtain the expression
\[
n\sum_{e=d/2-t}^{d-2t}p^e(1-p)^{n-e}\sum_{w=d}^{2(e+t)}\binom nw
\binom w{w/2-t}\binom{n-w}{e-w/2+t}.
\]
The remaining part of the analysis of this case
proceeds as above except that $t$ is replaced by $-t$
throughout. In particular, $\omega^\ast=2\rho(1-\rho)+2\tau(1-2\rho)$,
the optimum on $\rho$ is attained for $\rho_0^-$ and so on.
\qed

\providecommand{\bysame}{\leavevmode\hbox to3em{\hrulefill}\thinspace}
\providecommand{\MR}{\relax\ifhmode\unskip\space\fi MR }
\providecommand{\MRhref}[2]{%
  \href{http://www.ams.org/mathscinet-getitem?mr=#1}{#2}
}
\providecommand{\href}[2]{#2}


\begin{thebibliography}{10}


\bibitem{bar02b}
A.~Barg and G.~D. Forney, Jr., \emph{Random codes: {M}inimum distances and
  error exponents}, IEEE Trans. Inform. Theory \textbf{48} (2002), no.~9,
  2568--2573.

\bibitem{bla87}
R.~E. Blahut, \emph{Principles and practice of information theory},
  Addison-Wesley, Reading, MA, 1987.

\bibitem{blo82}
E.~L. Blokh and V.~V. Zyablov, \emph{Linear concatenated codes}, Nauka, Moscow,
  1982, (In Russian).

\bibitem{cox63}
H.~S.~M. Coxeter, \emph{An upper bound for the number of equal nonoverlapping
  spheres that can touch another of the same size}, Proc. Symp. Pure Math.,
  vol.~7, Providence: AMS, 1963, pp.~53--72.

\bibitem{csi81}
I.~Csisz{\'a}r and J.~K{\"o}rner, \emph{Information theory. {C}oding theorems
  for discrete memoryless channels}, Akad{\'e}miai Kiad{\'o}, Budapest, 1981.

\bibitem{bru58}
N.~G. de~Bruijn, \emph{Asymptotic methods in analysis}, North-Holland,
  Amsterdam, 1958.

\bibitem{for66}
G.~D. Forney, Jr., \emph{Concantenated codes}, MIT Press, Cambridge, MA, 1966.

\bibitem{for68b}
\bysame, \emph{Exponential error bounds for erasure, list, and decision
  feedback schemes}, IEEE Trans. Inform. Theory \textbf{14} (1968), no.~2,
  206--220.

\bibitem{gal63}
R.~G. Gallager, \emph{Low-density parity-check codes}, MIT Press, Cambridge,
  MA, 1963.

\bibitem{gal68}
\bysame, \emph{Information theory and reliable communication}, John Wiley \&
  Sons, New York e.a., 1968.

\bibitem{lev77}
V.~I. Levenshtein, \emph{Bounds on the probability of undetected error},
  Problemy Peredachi Informatsii \textbf{13} (1977), no.~1, 3--18.

\bibitem{olv97}
F.~W.~J. Olver, \emph{Asymptotics and special functions}, A K Peters Ltd.,
  Wellesley, MA, 1997. \MR{97i:41001}

\bibitem{pol94a}
G.~Sh. Poltyrev, \emph{Bounds on the decoding error probability of binary
  linear codes via their spectra}, IEEE Trans. Inform. Theory \textbf{40}
  (1994), no.~4, 1284--1292.

\bibitem{pol94b}
\bysame, \emph{On coding without restrictions for the {AWGN} channel}, IEEE
  Trans. Inform. Theory \textbf{40} (1994), no.~4, 409--417.

\bibitem{ran55}
R.~A. Rankin, \emph{The closest packing of spherical caps in $n$ dimensions},
  Proc. Glasgow Math. Assoc. \textbf{2} (1955), 139--144.

\bibitem{sha02}
S.~Shamai~(Shitz) and I.~Sason, \emph{Variations on the Gallager bounds,
  connections and applications}, IEEE Trans. Inform. Theory (2002), no.~12,
  3029--3051.

\bibitem{sha59}
C.~E. Shannon, \emph{Probability of error for optimal codes in a {G}aussian
  channel}, Bell Syst. Techn. Journ. \textbf{38} (1959), no.~3, 611--656.

\bibitem{shu99}
N.~Shulman and M.~Feder, \emph{Random coding techniques for nonrandom codes},
  IEEE Trans. Inform. Theory \textbf{45} (1999), 2101--2104.

\bibitem{vit79}
A.~J. Viterbi and J.~K. Omura, \emph{Principles of digital communication and
  coding}, McGraw-Hill, 1979.

\bibitem{wyn65}
A.~D. Wyner, \emph{Capabilities of bounded discrepancy decoding}, Bell Syst.
  Techn. Journ. (1965), 1061--1122.

\end{thebibliography}
\end{document}